# Amorphized graphene: a stiff material with low thermal conductivity

Bohayra Mortazavi[*, 1], Zheyong Fan[2, 3], Luiz Felipe C. Pereira[4], Ari Harju[3], Timon Rabczuk[1,5,#]

[1]*Institute of Structural Mechanics, Bauhaus-Universität Weimar, Marienstr. 15, D-99423 Weimar, Germany.*
[2]*School of Mathematics and Physics, Bohai University, Jinzhou, China.*
[3]*COMP Centre of Excellence, Department of Applied Physics, Aalto University, Helsinki, Finland.*
[4]*Departamento de Fisica Teorica e Experimental, Universidade Federal do Rio Grande do Norte, 59078-970 Natal, Brazil.*
[5]*School of Civil, Environmental and Architectural Engineering, Korea University, Seoul, South Korea.*

All-carbon heterostructures have been produced recently via focused ion beam patterning of single layer graphene. Amorphized graphene is similar to a graphene sheet in which some hexagons are replaced by a combination of pentagonal, heptagonal and octagonal rings. The present investigation provides a general view regarding phonon and load transfer along amorphous graphene. The developed models for the evaluation of mechanical and thermal conductivity properties yield accurate results for pristine graphene and acquired findings for amorphized graphene films are size independent. Our atomistic results show that amorphous graphene sheets could exhibit a remarkably high elastic modulus of ~500 GPa and tensile strengths of ~50 GPa at room temperature. However, our results show that mechanical properties of amorphous graphene decline at higher temperatures. Furthermore, we show that amorphized graphene present a low thermal conductivity ~15 W/mK which is two orders of magnitude smaller than pristine graphene, and we verify that its thermal conductivity is almost insensitive to temperature since it is dominated by phonon-defect scattering rather than phonon-phonon scattering. Finally, our results show that amorphized graphene structures present a remarkably high elastic modulus and mechanical strength, along with a low thermal conductivity, which is an unusual combination for carbon-based materials.

*Corresponding author (Bohayra Mortazavi): bohayra.mortazavi@gmail.com
Tel: +49 157 8037 8770,
Fax: +49 364 358 4511
#Timon.rabczuk@uni-weimar.de

## 1. Introduction

Carbon is arguably one of the most versatile chemical elements known. In recent decades the interest in this element has increased considerably due to the appearance of novel carbon allotropes such as nanotubes and graphene [1–4]. Carbon atoms are the building blocks of some of the hardest and toughest materials ever known. Furthermore, the thermal conductivity of carbon-based materials varies from very low to some of the highest known values [5–9]. Graphene, the planar form of carbon atoms with the honeycomb atomic lattice has been experimentally confirmed to yield exceptionally high mechanical [10] and thermal conduction properties [11–13] that outperform all known material. The connection between tensile rigidity and heat conductivity is very clear in these carbon-based materials: the ones with high elastic modulus also present high thermal conductivity. In the case of nanotubes, graphene and diamond the high thermal conductivity is mostly due to the strength of carbon-carbon bonds and the light weight of carbon atoms [14]. Nonetheless, under certain circumstances, this relationship can be weakened, and a material could be produced such that it possesses a high mechanical strength but a low thermal conductivity. As an example, such a combination is useful for thermoelectric application in which the thermal conductivity is preferred to be low [15,16] whereas the electron charge mobility should be at high level. Tailoring the properties of carbon based materials are commonly achieved by defect engineering [17–19] or chemical doping [20–23]. Interestingly, it was recently shown [17,18] that ion beam patterning of pristine graphene can be exploited for the fabrication of all-carbon single-layer heterostructures. These experimental advances raise the importance of further studies to explore the properties of these novel structures. Because of the complexities and cost of experimental characterization at atomic-scale, accurate theoretical calculations are currently considered as versatile and promising alternatives [24,25] to assess the properties of these novel advanced 2D materials. From experimental point of view, establishing relations between the defect concentration and the mechanical and heat conduction response of graphene not only ask for complicated characterization techniques but also more importantly require experimental fabrication of graphene sheets with controlled defect concentrations. These complexities can explain why up-to-date such relations have not been yet experimentally realized for defective graphene.

In this work we therefore study mechanical and heat transport properties of amorphous planar carbon structures, such as the ones recently produced by ion beam irradiation of pristine graphene [17]. We employ atomistic classical molecular dynamics simulations to probe mechanical and heat transport properties of amorphized graphene sheets with various concentrations of defects at different temperatures. In line with experimental tests, the amorphous structures in the present work were constructed by incorporation of defects. This way, our study provide a general viewpoint on the mechanical and thermal conductivity response of defective to amorphized graphene. Our results show that a highly amorphous graphene sheet could exhibit a remarkably high elastic modulus of ~500 GPa and tensile strengths of ~50 GPa at room temperature, while presenting a low thermal conductivity around 15 W/mK.

## 2. Molecular dynamics modeling

In order to construct atomistic models of amorphized graphene structures, we developed an algorithm in which the defect concentration of graphene sample could be varied. Stone-Wales defects are the most common type of defect observed in graphene which does not involve any removed or added atoms [19,26–28]. A single Stone-Wales defect can be formed by a 90 degree rotation of a C-C bond, in which four carbon hexagons are transformed into two pentagons and two heptagons. In our approach, Stone-Wales defects were randomly distributed in a pristine graphene sheet while preserving the periodicity of the structures [28]. In the present study, we define the defect concentration as the ratio of non-hexagonal rings in the amorphized structure with respect to the total number of hexagons in the initial pristine sheet. After creating the atomic positions for amorphized graphene structures with different defect concentrations, the structures were relaxed and equilibrated via molecular dynamics simulations. In all molecular dynamics simulations performed in this study, periodic boundary conditions were applied in the planar directions not only to remove the effect of free atoms on the boundaries but also to minimize finite size effects.

We have also used classical molecular dynamics simulations to evaluate mechanical and thermal transport properties of amorphized graphene. The accuracy of predictions based on molecular dynamics simulations are dependent on the selection of accurate interatomic potential functions to describe atomic interactions. Here, the interaction between carbon atoms is modeled by the Tersoff potential [29,30] with a parameter set optimized by Lindsay and Broido [31]. This optimized Tersoff potential predict phonon dispersion curves of graphite in close agreement with experiments [31]. Among all available force-fields for molecular dynamics modeling of graphene, recent simulations indicate that the optimized

Tersoff potential is an accurate and computationally efficient choice to model both thermal conductivity [5,14,32,33] and mechanical [28] response of graphene.

Mechanical properties of amorphized graphene structures were evaluated by performing uniaxial tensile tests at different temperatures: 300 K, 500 K and 700 K. These simulations were carried out with LAMMPS [34]. The time step of simulations for the uniaxial deformations was fixed at 0.25 fs. Prior to application of uniaxial loading conditions, all samples were relaxed and equilibrated using Nosé-Hoover barostat and thermostat method (NPT) with damping parameters of 2.5 fs and 250 fs for temperature and pressure, respectively. During loading, the periodic size of the simulation box along the loading direction was increased at every time step by a constant engineering strain rate of $2\times10^8$ s$^{-1}$. In order to guarantee uniaxial stress condition during loading, the pressure along the perpendicular direction was coupled to a barostat in order to keep it around zero, on average. The NPT method also controls the temperature fluctuations during loading. The engineering strain at each time step was calculated by multiplying the total time of uniaxial loading by the applied engineering strain rate. We then calculated the engineering Virial stresses at every strain level and averaged them during every 250 fs intervals to obtain smooth engineering stress-strain curves. In the stress calculations, the thickness of amorphized graphene membranes was set at 0.335 nm.

The thermal conductivity of pristine and amorphized graphene sheets has been calculated via equilibrium molecular dynamics simulations (EMD) with periodic boundary conditions in the graphene plane. For each defect concentration and temperature, the simulation cell is relaxed to achieve zero stress along in-plane directions, and the equations of motion are integrated with a 0.25 fs timestep.

The heat flux vector is calculated with the appropriate form for many-body potentials, including a kinetic part and a potential part as described in [14] via:

$$J = J_{kin} + J_{pot} = \sum_i v_i E_i + \sum_i \sum_{j \neq i} (r_j - r_i) \left( \frac{\partial U_j}{\partial r_{ji}} \cdot v_i \right) \quad (1)$$

where $v_i$ is the velocity of atom i, $r_i$ is the position vector of atom i, $E_i$ is the energy of atom i, $r_{ji}$ is the position vector from atoms j to i, and $U_j$ is the potential energy associated with atom j.

The components of the thermal conductivity tensor are obtained from MD trajectories as:

$$k_{\alpha\beta} = \frac{1}{V k_B T^2} \int_0^\infty \langle J_\alpha(0) J_\beta(t) \rangle \, dt \quad (2)$$

where $k_B$ is Boltzmann's constant, T is the simulation temperature, and V is the volume of the graphene sheet defined as the surface area times a nominal thickness of 0.335 nm. Thermal transport in pristine and amorphized graphene sheets is isotropic in the sheet plane, such that the thermal conductivity for a given sample is given by the average of the x- and y-components. Furthermore, in order to achieve statistical accuracy in our results, each reported value of κ is obtained by averaging over several independent simulations. We also note that the production time for each independent simulation required for achieving a given relative statistical accuracy depends crucially on the magnitude of the computed thermal conductivity: this way for pristine graphene this time is 100 ns, while for amorphized graphene, it is two orders of magnitude smaller. Worthy to remind that our recent investigation [14] demonstrated that the LAMMPS implementation of EMD method underestimates the thermal conductivity of graphene due to the use of an inaccurate heat flux formula for many-body potentials. In this study we therefore used an efficient code (implemented on graphics processing units) which implements the accurate heat flux formula as shown in Eq. 1 to compute the thermal conductivity.

In this study, we developed relatively large structures of amorphized graphene sheets consisting of 92,800 atoms (50 nm×50 nm). A sample with 35% defect concentration, relaxed and equilibrated with optimized Tersoff potential at room temperature is illustrated in Fig. 1. As a first finding, one can observe remarkable wrinkling and rippling of amorphized graphene sheets due to the presence of non-hexagonal rings in the structure [35]. Such an observation has been previously reported along graphene grain boundaries due to the existence of pentagon-heptagon pairs [35–38]. Our initial simulations also shown that the Tersoff potential could successfully stabilize defect regions throughout the sample. We note that the use of a small time step equal to 0.25 fs was necessary to avoid any instabilities during the evaluation of thermal and mechanical properties of amorphized graphene films. As it can be observed in the inset of Fig. 1, our amorphized models consist mainly of pentagon and heptagon rings and the formation of octagons is less probable. This observation is in agreement with experimental findings regarding amorphized graphene [17] in which it was revealed that in amorphous graphene the concentration of pentagon and heptagon rings is higher than the octagonal rings. Nonetheless, it is worthy to note that in the construction of amorphized structures, the concentration of octagonal rings could be increased by randomly distributing both Stone-Wales rotations and double-vacancy defects in the graphene sheet. However, we found that because of the limited reactivity of the optimized Tersoff potential it faces stability issues for structures with high concentration of octagonal rings.

**3. Results and discussions**

3.1 Mechanical properties

The calculated stress-strain response of pristine graphene at room temperature using the original set of parameters proposed by Lindsay and Broido [31] is illustrated in Fig. 2. In the Tersoff potential for carbon [30,31], a cutoff function is used for the covalent interaction for atom distances between 0.18 nm to 0.21 nm. However, this cutoff induces unphysically high fracture stresses during the tensile deformation when the C-C bonds are stretched longer than 0.18 nm. This is apparent from the stress-strain response for strain levels higher than 0.2 in which an unphysical strain hardening in the stress values could be observed. It should be noted that such an observation has also been reported with the AIREBO potential[39] in the simulation of tensile deformation of graphene [40,41]. To the best of our knowledge the AIREBO forcefield has been the mostly used potential for simulating mechanical properties of graphene. Regarding the AIREBO potential, the problem concerning the unphysical high fracture stresses was successfully solved by changing the cutoff value [40,41]. In this regard, the cutoff value of the AIREBO potential was increased from 0.17 nm to 0.2 nm which was shown to yield more accurate predictions for the mechanical strength of graphene in comparison with the original cutoff [41]. Consequently, we also modified the cutoff of Tersoff potential from 0.18 nm to 0.20 nm and we plot the obtained stress-strain response of pristine graphene in Fig. 2. Our calculated stress-strain curves for original and modified optimized Tersoff potential show that the cutoff modification does not affect the stress-strain values for strain levels smaller than 0.17 but it successfully removes the unphysical strain hardening at higher strain levels. The elastic modulus is calculated using the slope of the initial linear region of the stress–strain curves. In this way, our molecular dynamics model predict an elastic modulus of 960±10 GPa and tensile strength of 132 GPa for defect-free graphene. These predictions are in excellent agreements with experimental results of 1000±100 GPa for the elastic modulus and tensile strength of 130±10 GPa for pristine graphene [10]. It should be noted that using the AIREBO potential a tensile strength of 100-125 GPa at failure strains of 0.13-0.20 were predicted for pristine graphene [40,41]. Taking into consideration that the Tersoff potential is considerably faster than AIREBO potential in terms of computational costs along with its more accurate predictions for mechanical properties for graphene, we conclude that the modified optimized Tersoff potential is the most appropriate choice for the evaluation of mechanical properties of carbon-based structures. Further calculations for pristine graphene at higher temperatures reveal that the elastic modulus of graphene decreases slightly whereas the decline in tensile strength is more considerable.

Using the modified optimized Tersoff potential we found that defect-free graphene extends uniformly during uniaxial tensile loading. In this case we observed that the specimen maintains its pristine structure up to the tensile strength point. The tensile strength was found to be a place in which the first debonding occurs between two adjacent carbon atoms. This bond breakage results in the formation of cracks that rapidly grow and finally lead to the sample rupture. For pristine graphene, the initial void formation and subsequent failure occur at very close strain levels suggesting a brittle failure mechanism. In Fig. 3, the deformation process of an amorphized graphene sheet with 35% concentration of defects at various stages of loading is depicted. As discussed earlier, the relaxed structure presents remarkable wrinkling and rippling due to the presence of defects (Fig. 3a). However, during loading the structure is flattened along the loading direction. Nevertheless, due to the presence of defects the structure still presents a waviness pattern perpendicular to the loading direction (Fig. 3b). During the initial stages of uniaxial tensile loading we could observe the initiation of C-C bonds breakage. In this case, the initial debondings were found to be more favorable to occur along the octagonal rings. Nonetheless, by increasing strain levels, more bonds are broken (Fig. 3b) forming larger voids that are visible throughout the entire sample. The tensile strength is then found to take place at the point in which the coalescence of existing voids occurs (Fig. 3c) forming a large crack almost perpendicular to the loading direction. The crack formation in the sample results in a characteristic sharp decrease in the stress-strain response. Our results suggest that during crack growth in amorphized graphene, mono-atomic carbon chains form between two sides of the crack resisting against the crack growth. In addition to the formation of mono-atomic carbon chains, the existence of an arbitrary distribution of defects in front of the crack tip force the crack growth to deviate from a straight direction resulting in the formation of rough and irregular edges.

In Fig. 4, the room-temperature stress-strain response of amorphized graphene sheets with different defect concentrations are illustrated. The results for a particular defect concentration are shown for two independent samples with different defect distributions. As it can be observed, there exist clear differences in loading behavior between stress-strain curves for different defect concentrations. In addition, it is shown that the stress-strain responses for a given defect concentration are in good agreement up to the tensile strength. This confirms that the simulated samples were large enough in order to be representative of a volume element of a real system. Nonetheless, the tensile strength for two samples with the same defect concentration but independent defect configurations are found to be different. As discussed earlier, the characteristic sign of the tensile strength is the coalescence of initial voids that were gradually developed during sample deformation. Therefore, the tensile strength and the corresponding failure strain present a stochastic nature specially when temperature effects are considered in the model. Unlike the results for pristine graphene, the stress-strain curves for amorphized graphene sheets present a non-linear response at low strain levels. As illustrated in Fig. 4, this initial non-linear response region increases by increasing the defect concentration. It is important to remind that by increasing the defect concentration the structures undergo severe rippling and wrinkling because of the non-hexagonal structure resulting in the contraction of the sheet. This way, at initial strain levels the applied stress flattens the sheets (Fig. 3b), therefore the stress values increase gradually and in a non-linear pattern. Accordingly, in the evaluation of elastic modulus, we neglected this initial non-linear regime and considered only the subsequent linear response in stress-values to report the elastic modulus.

In Fig. 5, the calculated elastic modulus and tensile strength of amorphized graphene films as a function of defect concentration at different temperatures are presented. In all studied cases clear decreasing trends in elastic modulus and tensile strength exist by increasing defect concentration or temperature. Interestingly, our results show that highly defective amorphized graphene could exhibit a remarkably high elastic modulus of around 500 GPa and tensile strengths of around 50 GPa at room temperature. Therefore, we predict that highly defective amorphized graphene structures keep their load bearing abilities under high tensile stresses by almost two orders of magnitude higher than high strength steel and titanium alloys. In addition, we found that at even at elevated loading temperatures the mechanical properties of amorphized graphene are still at considerably high levels.

3.2 Thermal conductivity

Next, we study the thermal conductivity of amorphized graphene samples. We begin by evaluating the thermal conductivity of pristine graphene in order to have a benchmark for comparison. In Fig. 6, the calculated thermal conductivities of pristine graphene as a function of correlation time at temperatures of 300 K and 500 K are depicted. In order to obtain converged thermal conductivities using the EMD method, we performed several simulations (each simulation lasts 100 ns in the production stage) with uncorrelated initial conditions. For each independent simulation, the results were averaged along the planar directions (x and y directions). The converged thermal conductivity was finally evaluated by averaging the results over the several independent simulations (as shown in Fig. 6). The results in Fig. 6 predict a thermal conductivity of 2700±80 W/mK for pristine graphene at 300 K which converges at a correlation time of around 500 ps. This value is also within the reported range of 2600-3050 W/mK [5,32,42] for thermal conductivity of pristine graphene obtained by the non-equilibrium molecular dynamics simulations based on the optimized Tersoff potential. The thermal conductivity of pristine graphene at 500 K converges to 1500±45 W/mK at an earlier correlation time around 300 ps, because of higher phonon-phonon scattering rates at higher temperatures. We would like to note that in comparison with non-equilibrium molecular dynamics (NEMD) method for the evaluation of thermal conductivity, the EMD method is much less sensitive to the sample size [43]. In the NEMD method the fixed boundaries result in the significant size dependency so that by increasing the graphene size the thermal conductivity increases [5,32]. Interestingly, using the EMD method our recent study [14] confirms that relatively small samples with around $10^4$ atoms can accurately predict the thermal conductivity of graphene, due to the absence of boundaries. Nevertheless, we constructed samples with almost $10^5$ atoms which fully guarantee that our findings for amorphized graphene thermal conductivities are independent of sample size.

In Fig. 7, we plot the calculated thermal conductivities as a function of correlation time for two amorphized graphene samples with defect concentrations of 5% and 35 % at 300 K and 500 K temperatures. We observe that in amorphized films the thermal conductivity converges at correlation times more than two orders of magnitude shorter than those for pristine samples. Similarly to our simulations for mechanical properties, for each defect concentration we constructed two samples with different random defect configurations. The results for each sample agree within error bars, which show that averaging over 10 simulations with independent initial configurations is enough to obtain representative results for the conductivity. It is worth of notice that employing the non-equilibrium molecular dynamics method to calculate the thermal conductivity leads to instabilities in the simulations due to the imposed heat flux forces. We found that NEMD simulations of amorphous graphene structures with defect concentrations higher than 10% were unstable even at room temperature. Therefore, in this work we employ the equilibrium molecular dynamics method proposed by Fan *et al.*[14] which is computationally fast, stable and accurate for the evaluation of thermal conductivity of various systems.

Calculated thermal conductivities of amorphized graphene samples with different defect concentrations at 300 K and 500 K are shown in Fig. 8. Our results reveal a drastic decline in the thermal conductivity of amorphous graphene by increasing defect concentration. Amorphized graphene sheets present a thermal conductivity almost two orders of magnitude smaller than pristine films. Interestingly, the temperature is found to play a minor role on the thermal conductivity of amorphized graphene. For pristine samples the thermal conductivity is inversely proportional to the temperature, as expected when phonon-phonon scattering plays the major contribution toward thermal resistance [33,44]. However, by increasing defect concentration in amorphous graphene films the effective thermal resistance becomes less and less sensitive to temperature effects. Our results predict that thermal conductivity of amorphized graphene films with defects concentrations higher than 20% are insensitive to temperature. This finding shows that phonon-defect scattering is the main factor dominating heat transport in amorphous graphene films.

To better understand the underlying mechanism responsible for the reduction in thermal conductivity of amorphous graphene in comparison with pristine graphene we calculated the phonon density of states (DOS). The DOS were obtained by post-processing 100 ps trajectories in which atomic velocities were recorded every 2.5 fs. The DOS was computed by calculating the Fourier transform of the velocity autocorrelation function, such that:

$$\text{DOS }(\omega) = \int_0^\infty \frac{\langle \boldsymbol{v}(t)\boldsymbol{v}(0)\rangle}{\langle \boldsymbol{v}(0)\boldsymbol{v}(0)\rangle} e^{-i\omega t} dt \qquad (3)$$

where ω is the frequency and the *v* is the atomic velocity. The calculated DOS for pristine and amorphous graphene films are presented in Fig. 9. As the defect concentration increases from 0%(pristine) to 12.5% and 35% we observe a broadening of most peaks, and considerable damping of the optical modes around 50 THz. Since phonon lifetimes are inversely proportional to the width of such peaks [45,46], and the thermal conductivity is proportional to phonon lifetimes [8,45,46], the broadening observed in DOS as defect concentration increases is associated with a reduction in conductivity, consistent with data in Figs. 7 and 8. Furthermore, in Fig. 9 we also observe a reduction in DOS of phonon modes with frequency below 4 THz. These long-wavelength acoustic phonons are the major heat carriers in graphene and related materials [8,47]. Therefore, the observed reduction in DOS of low-frequency modes is also consistent with the decrease in thermal conductivity for amorphized samples, similar to what was recently observed for polycrystalline graphene and boron nitride [44,48].

In order to acquire more information regarding the thermal conductivity reduction in amorphized graphene, we obtained the phonon dispersion relations within the harmonic approximation, for pristine and amorphized samples, and from these we calculated the group velocities. The group velocities are given by:

$$v_g = \frac{d\omega}{dq} \qquad (4)$$

where ω is the frequency of a given mode and *q* stands for the wave vector. In Fig. 10 we present the absolute value of the group velocities as a function of frequency for a pristine sample and an amorphized sample with the same number of atoms and same approximate dimensions. The data shows a large reduction in phonon group velocities in the amorphized sample relative to defect-free graphene. Since the thermal conductivity is proportional to the square of the group velocities (see for example, Eq. 10 in Ref. [7]), we find further evidence to justify the reduction in thermal conductivity observed in amorphized graphene. Worthy to note that defects in graphene can also affect the charge mobility and its electronics properties as well [49–51].

To further probe which type of defect contribute more to the heat resistance along the amorphized graphene sheets, we also calculated the phonon group velocities for two samples made by including the octagon-pentagon rings or heptagon-pentagon rings. The results depicted in Fig. 11 clearly confirm that the phonon group velocities are lower along the amorphized graphene made by octagon-pentagon rings in comparison with the one constructed using the heptagon-pentagon. This effect is more noticeable for the low frequency phonons, which play major role for the thermal conductivity of graphene. This way, an amorphized graphene with higher percentage of octagons will present a lower thermal conductivity. Nevertheless, experimental observation for amorphized graphene [17] and graphene grain boundaries [35,36] reveal limited concentration of octagons which implies that heptagons rings are thermodynamically more stable and favorable for carbon atoms. Higher stability of heptagons in comparison with octagons also well agree with our phonon calculations which reveal stronger and faster phonon transfer through heptagons.

Our classical atomistic simulations reveal that a highly amorphized graphene yields a thermal conductivity around 15 W/mK. This thermal conductivity is by two orders of magnitude smaller than those for pristine graphene [8,9,11,12] and isotopically modified graphene [52] too. In addition, amorphized graphene yields a thermal conductivity by an order of magnitude smaller than nanocrystalline graphene [53,48]. Worthy to mention that the thermal conductivity of amorphized graphene is also by several times smaller than the thermal conductivity of graphene laminates [32,54,55] in which the heat transfer is strongly dependent on the weak van-der-waals forces acting between the contacting graphene membranes. As a matter of fact, the thermal conductivity of amorphized graphene is yet by an order of magnitude higher than the one for diamond-like carbon [8,56]. This comparison clearly highlights the outstanding and exceptional variable and tunable heat conduction response of carbon based structures.

## 4. Conclusion

We performed extensive classical molecular dynamics simulations to provide a general viewpoint concerning the mechanical response and thermal conductivity of amorphized graphene films. We developed large atomistic models of amorphized graphene films with defect concentrations ranging from 5% to 35%. Modifying the cutoff of the optimized Tersoff potential from 0.18 nm to 0.2 nm, we obtained mechanical properties for pristine graphene in a significant agreement with experimental results. Amorphized graphene samples were found to present remarkable wrinkling and rippling due to the presence of non-hexagonal rings. Investigating tensile deformation of amorphous graphene include we observed void creation at initial stress level and void extensions during uniaxial loading which lead to formation of cracks almost perpendicular to the loading direction. The formed cracks rapidly propagate by breaking mono-atomic carbon chains and leading to sample rupture. Amorphized graphene is therefore predicted to present ductile failure. Our atomistic modeling suggests that a highly amorphous graphene sample can exhibit a remarkably high elastic modulus of around 500 GPa and tensile strengths of around 50 GPa at room temperature. We also predict that amorphized graphene sheets present a thermal conductivity around 15 W/mK which is two orders of magnitude smaller than pristine graphene, and that temperature plays a minor role on the thermal conductivity of amorphized samples since phonon-defect scattering is the main source of thermal resistance in these defective samples. Finally, our results show that although amorphized graphene structures present a remarkably high elastic modulus and mechanical strength, their thermal conductivity is low, which is an unusual combination for carbon-based materials.

**Acknowledgment**

BM and TR greatly acknowledge the financial support by European Research Council for COMBAT project. ZF and AH are supported by the Academy of Finland through its Centres of Excellence Program (project no. 251748) and they acknowledge the computational resources provided by Aalto Science-IT project and Finland's IT Center for Science (CSC). LFCP acknowledges financial support from Brazilian government agency CAPES via its Science Without Borders program.

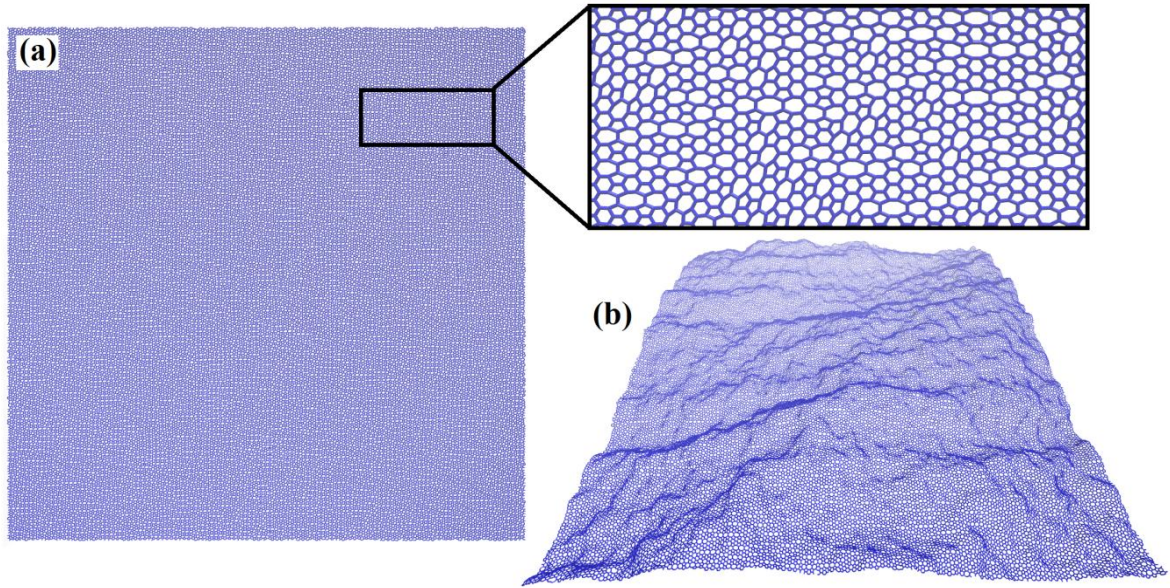

Fig. 1- (a) Atomistic model of a periodic amorphized graphene with 35% defect concentration made from 92,800 carbon atoms. The inset shows a detailed view focusing on a highly defective zone which shows that the structures is consisting of pentagonal, hexagonal, heptagonal and octagonal carbon rings that are randomly and irregularly distributed along the sheet. (b) The side view of the same structure after relaxation and equilibration at room temperature using the optimized Tersoff potential. VMD software is used for the illustration of structures [57].

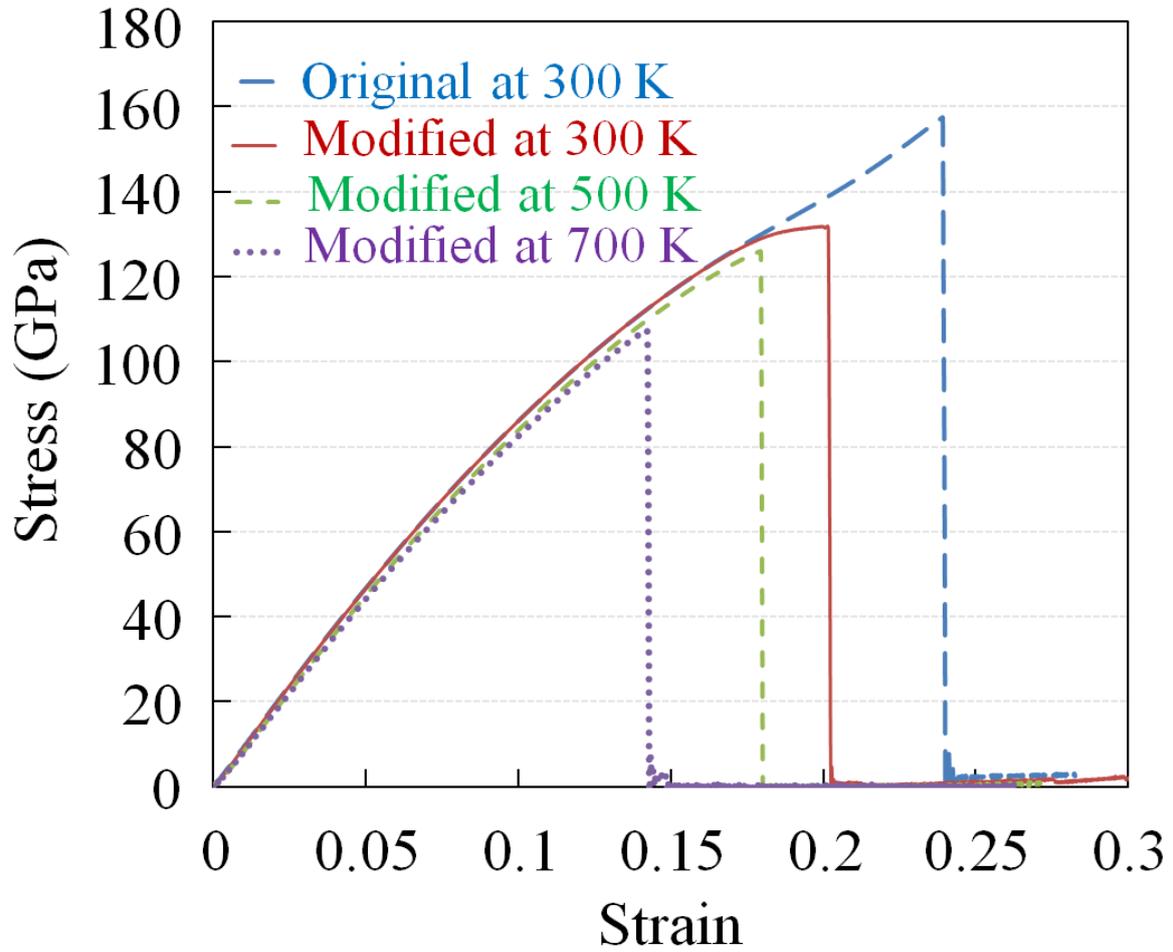

Fig. 2- Calculated uniaxial stress-strain response of defect-free pristine graphene using the optimized Tersoff potential by Lindsay and Broido [14]. At room temperature stress-strain curves are plotted using the original and modified optimized Tersoff potential. In the modified version the initial cutoff of the Tersoff potential was changed from 0.18 nm to 0.2 nm which yields more accurate results.

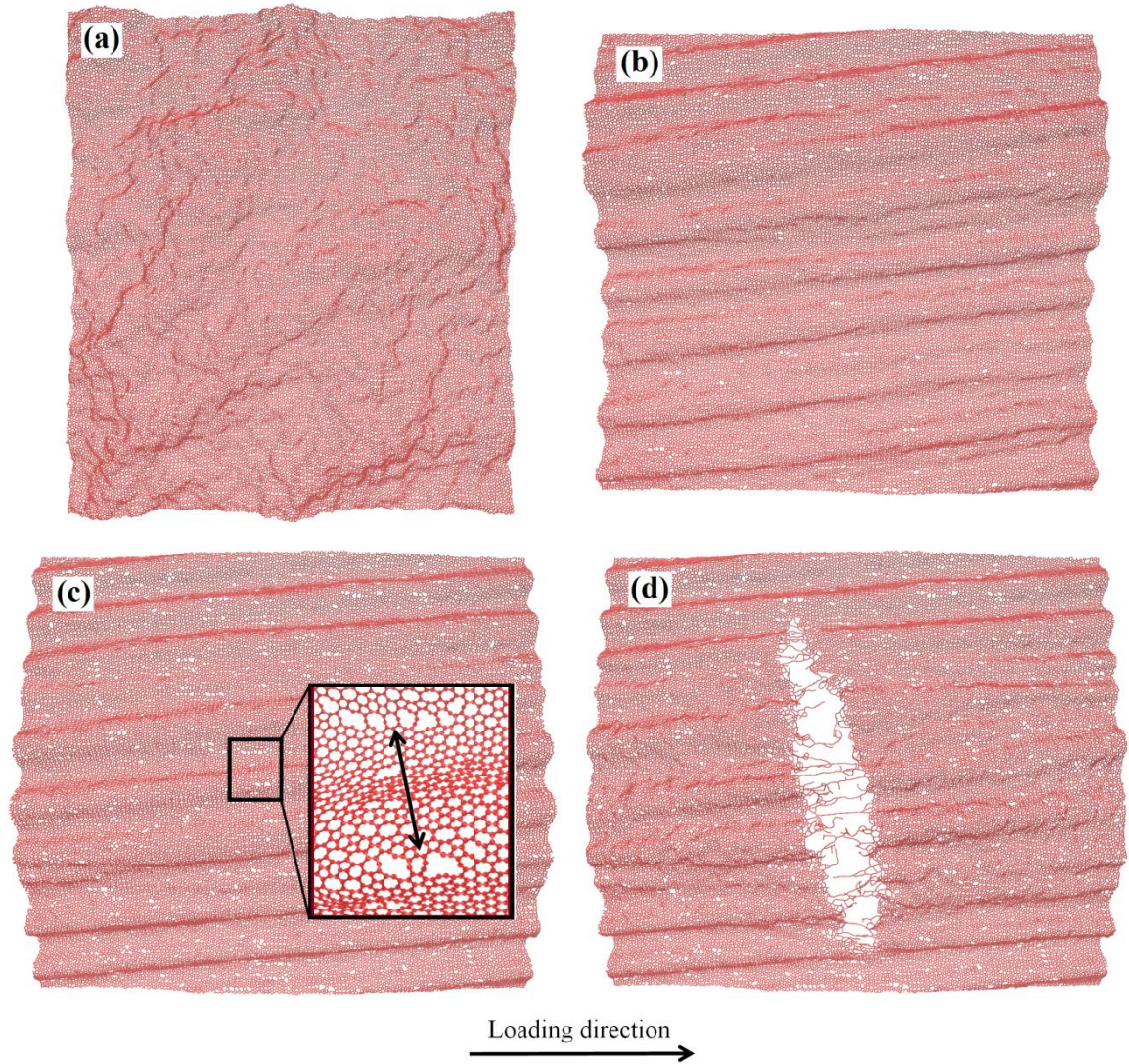

Fig. 3- Deformation process of an amorphized graphene sheet with 35% defect concentration under uniaxial tensile loading at different strain levels. (a) relaxed structure before loading, (b) structure at a strain of 0.12 contains extended defects throughout the sheet, (c) tensile strength point is found to be a time in which the crack coalescence occurs, (d) shortly after the tensile strength the crack extend in the sheet leading to sample rupture.

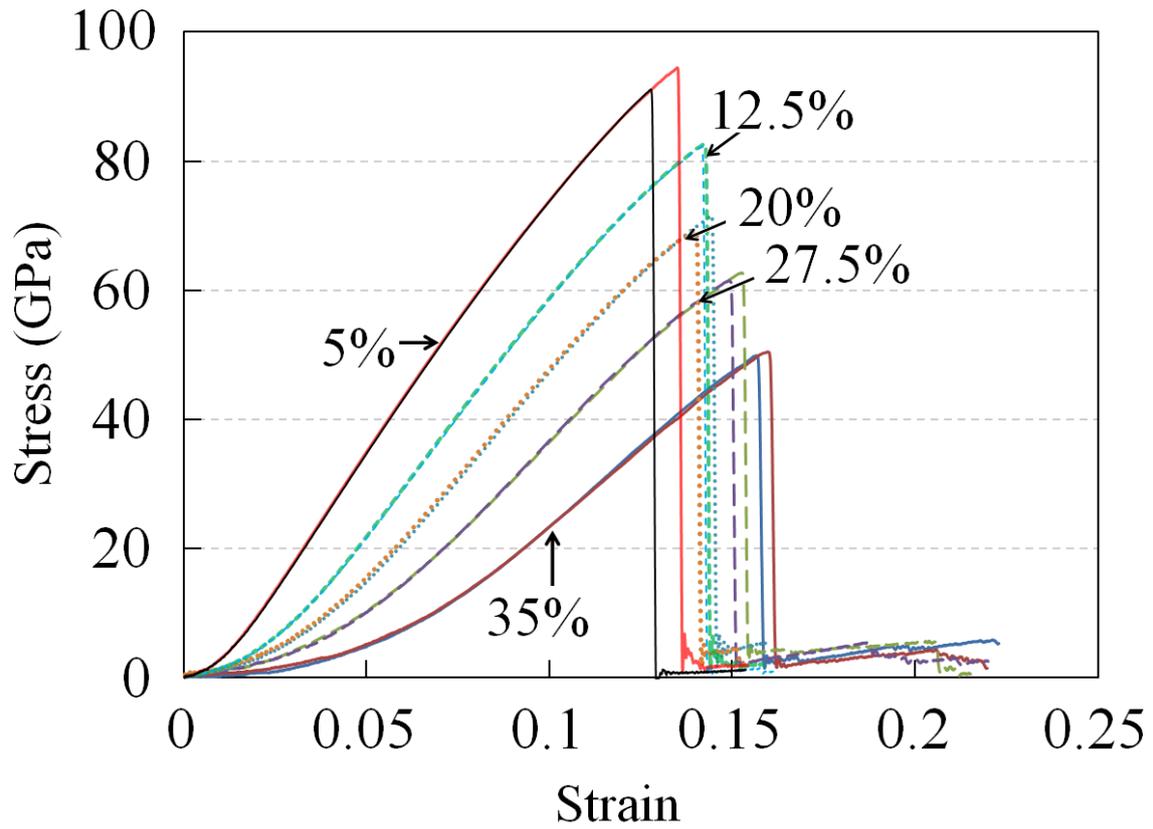

Fig. 4- Acquired stress-strain response of amorphous graphene sheets with different defect concentrations of 5% to 35% at room temperature. For each defect concentration, the results are plotted for two samples with different defect configurations.

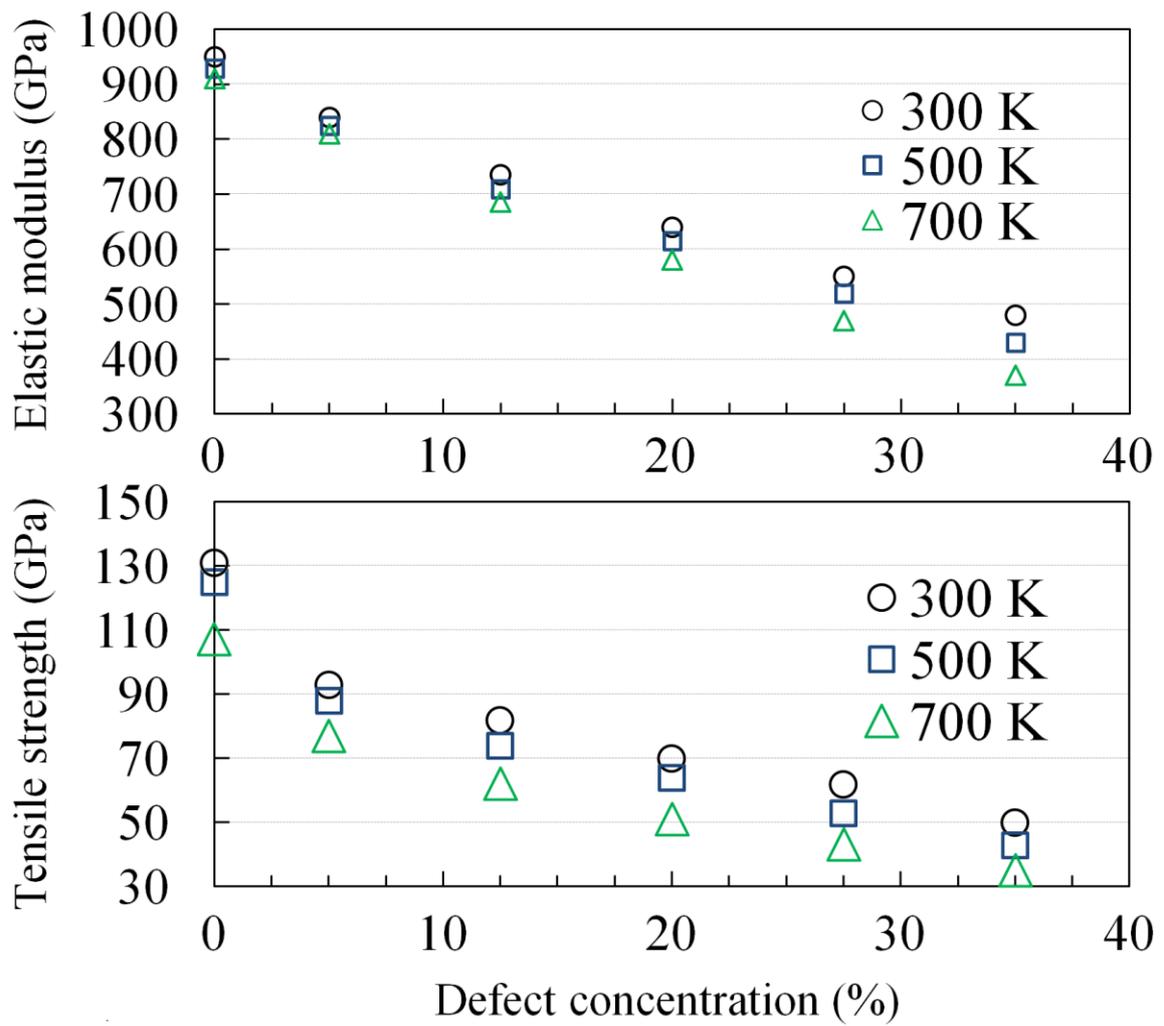

Fig. 5- Elastic modulus and tensile strength of different amorphized graphene sheets with different defect concentrations at various loading temperatures.

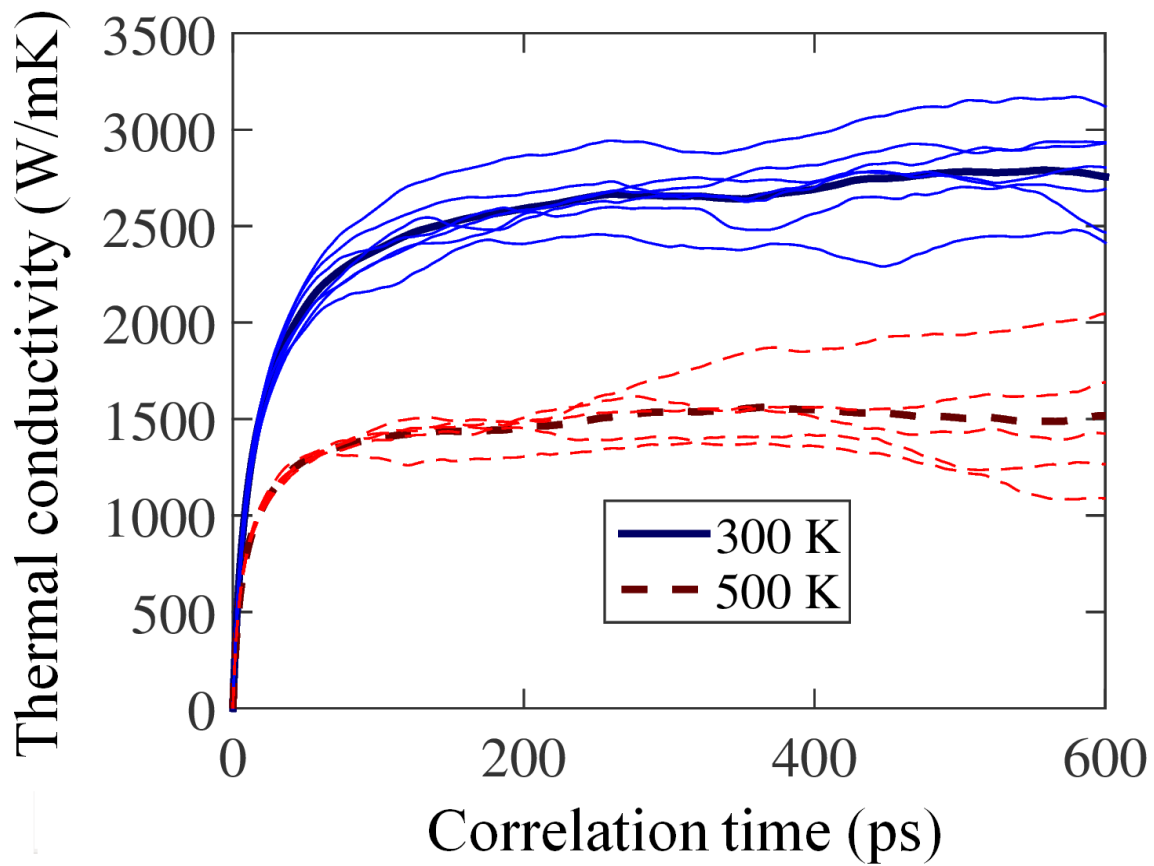

Fig. 6- Calculated thermal conductivities as a function of correlation time for pristine graphene at 300 K and 500 K. The thinner lines represent the results of independent simulations with different initial velocities and the thick lines show the ensemble average over the independent simulations.

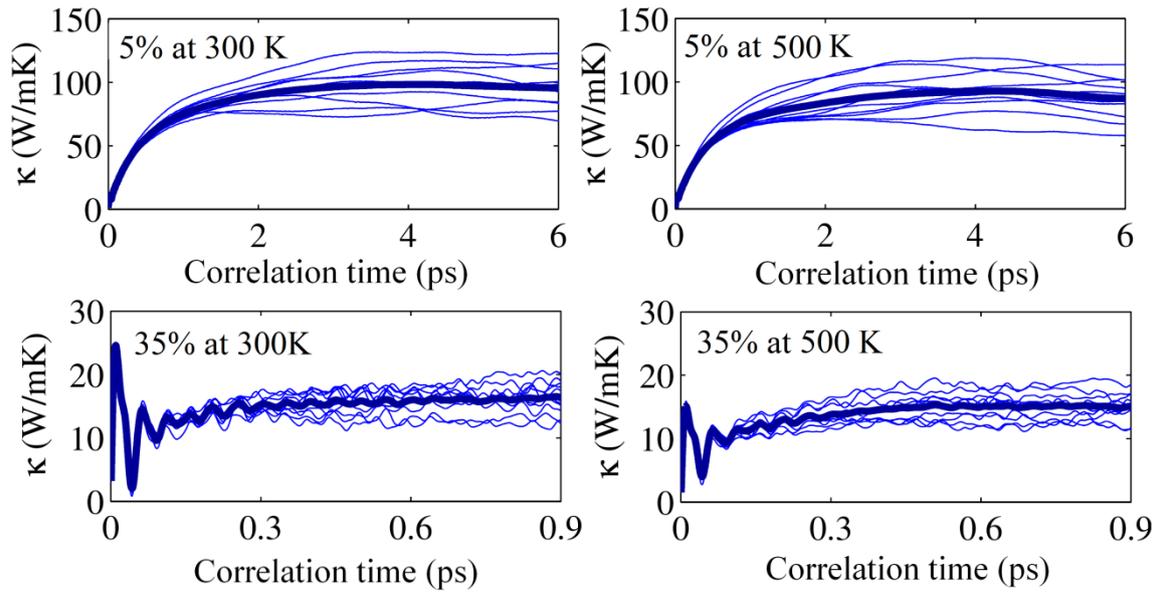

Fig. 7- Typical results of calculated thermal conductivities as a function of correlation time for single-layer amorphized graphene samples.

.

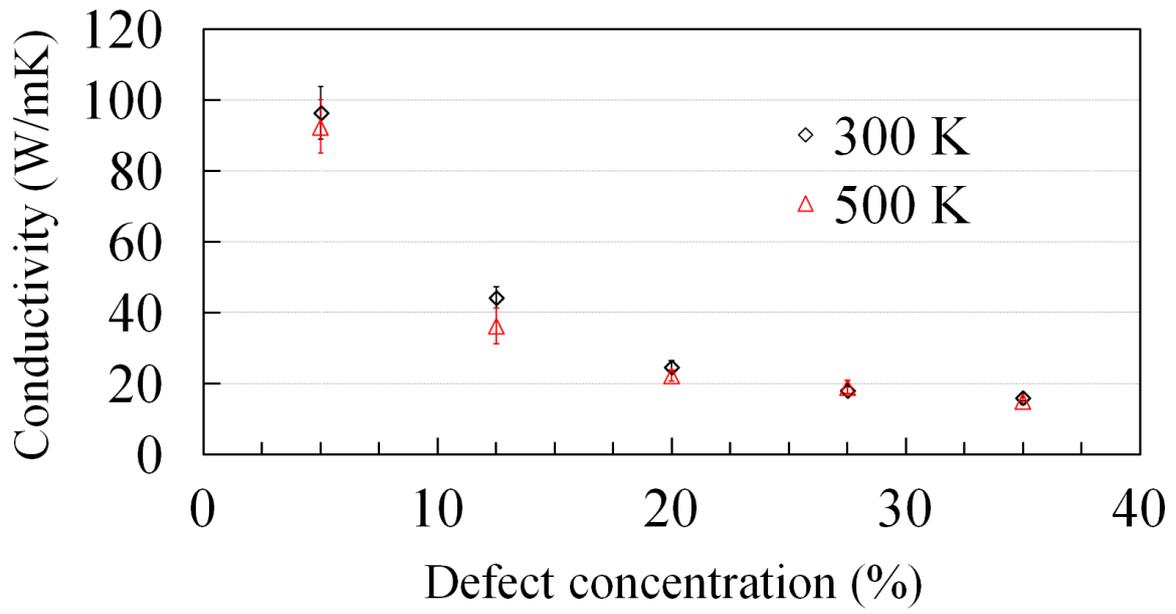

Fig. 8- Thermal conductivity of amorphized graphene films for different defect concentrations at 300 K and 500 K.

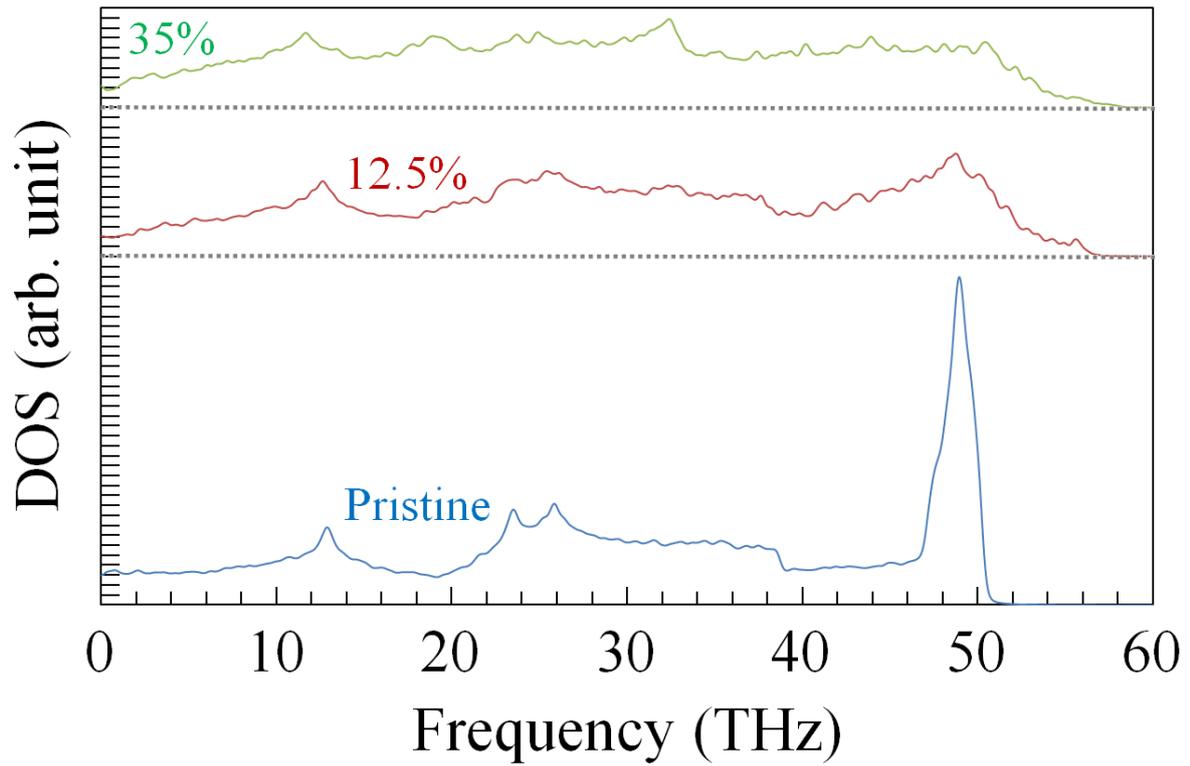

Fig. 9- Calculated vibrational density of states for pristine and two amorphized graphene films. In comparison with pristine graphene, the optical mode around 50 THz is damped considerably in amorphized graphene films. In addition, the population of acoustic modes with frequencies below 4 THz also decrease.

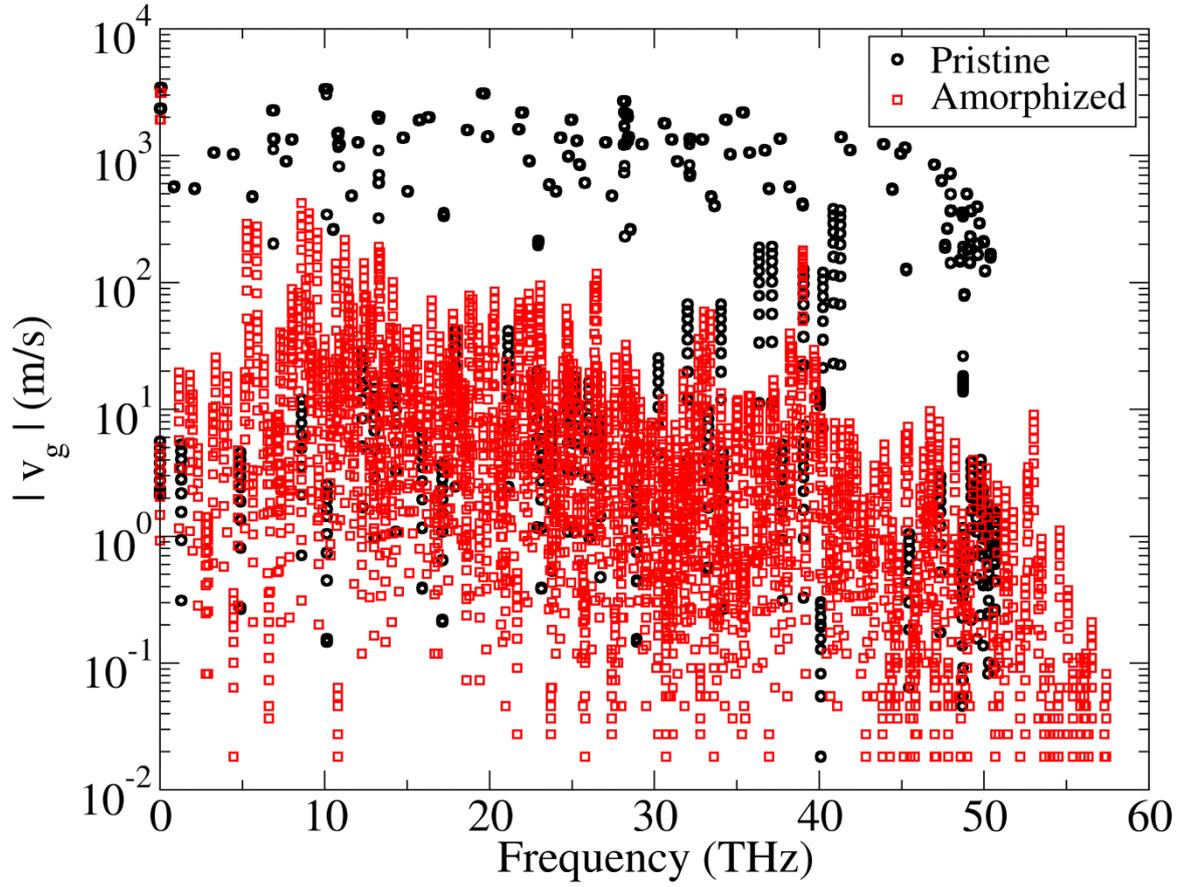

Fig. 10- Calculated phonon group velocities for pristine and amorphized graphene samples. In comparison with pristine graphene, the group velocities are considerably lower in amorphized graphene. This observation further corroborates the reduction observed in amorphized graphene samples.

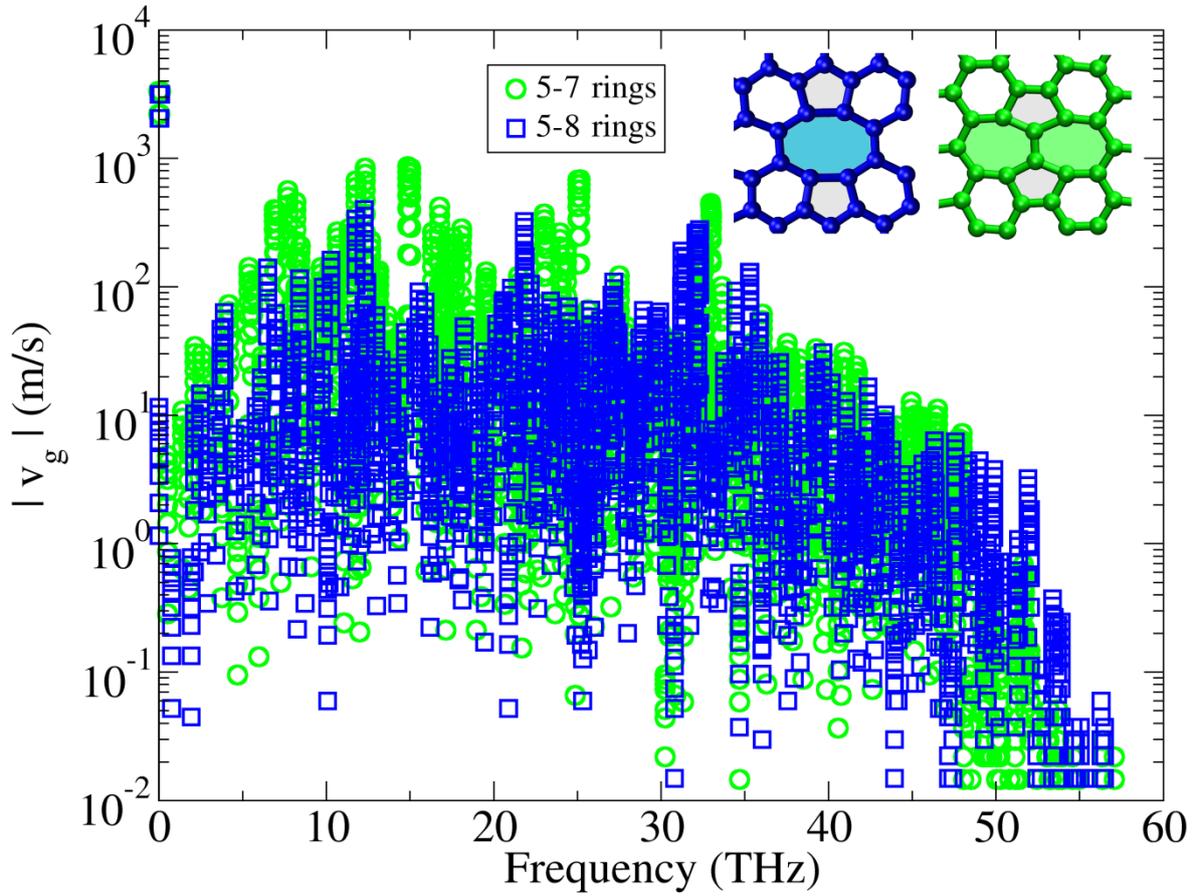

Fig. 11- Comparison of phonon group velocities for two amorphized graphene samples constructed by using 5-7 rings or 5-8 rings. The defect concentration for both samples is 15%. The insets show atomic configurations for the 5-7 and 5-8 rings.